\documentclass{camera}
\usepackage{natbib}
\usepackage{graphicx}  
\newcommand{\ls}{LS~5039}
\newcommand{\lsi}{LS~I~+61$^{\rm o}$303}
\newcommand{\psrb}{PSR~B1259-63}
\def\spose#1{\hbox to 0pt{#1\hss}}
\def\la{\mathrel{\spose{\lower 3pt\hbox{$\mathchar"218$}}
    \raise 2.0pt\hbox{$\mathchar"13C$}}}
\def\ga{\mathrel{\spose{\lower 3pt\hbox{$\mathchar"218$}}
    \raise 2.0pt\hbox{$\mathchar"13E$}}}

\begin{document}

%
\title{The $\gamma$-ray binaries \ls, \\ \lsi\ and \psrb}

%
\author{Guillaume Dubus}

%
\organization{Laboratoire Leprince-Ringuet, UMR 7638 CNRS, Palaiseau, France \\
Institut d'Astrophysique de Paris, UMR 7095 CNRS, Paris, France
}

\maketitle

\begin{abstract}
Three binaries are now established sources of emission at very high energies ($>$10$^{11}$~eV). They are composed of a massive star and a compact object. The emission can be due to the interaction of the relativistic wind from a young ms pulsar with the stellar wind of the companion \citep{maraschi}, by which rotation-power ends up as non-thermal flux. Variations at VHE energies are explained as due to $\gamma\gamma$ absorption and/or changes in shock location along the orbit. Resolved radio emission is due to cooling particles trailing the pulsar.
\end{abstract}

%

\section{The observational backdrop}
Thanks to the new generation of Cherenkov telescopes HESS and MAGIC, the past two years have seen a spectacular increase in the number of known sources emitting photons with energies greater than a TeV (see the contributions by Cui, Santangelo and Teshima in this volume). There are now about 40 TeV sources, mostly concentrated in the Galactic plane. Three of the Galactic sources have been associated with binary stars on the basis of their positional coincidence, variability and spectra: \psrb\ \citep{psrb}, \ls\ \citep{lspaper} and \lsi\ \citep{lsipaper}. 

\subsection{The binaries}
Those three {\em gamma-ray binaries} are composed of a massive star (O or Be type) and a compact object, neutron star or black hole, in an eccentric orbit. The orbital periods are $\approx 4$~days (\ls), 26~days (\lsi) and 1237~days (\psrb). Given the small mass functions, radial velocity measurements have been unable to pin down the nature of the compact object in \ls\ and \lsi. In both cases, a black hole would imply seeing the systems at $i\approx 30^{\rm o}$, while a neutron star would imply a more edge-on $i\approx 60^{\rm o}$. Regardless, the presence of a luminous early-type star, providing a copious source of seed photons for inverse Compton scattering, is probably instrumental to very high energy (VHE) $\gamma$-ray emission.

With distances ranging from 1.5-3~kpc, the TeV emission is at a level $\sim 10^{33-34}$~erg~s$^{-1}$, comparable to the X-ray emission from these sources. Remarkably, all have been detected in the radio, a feature otherwise shared by only a handful of high-mass X-ray binaries. Bar from periodic radio outbursts in \lsi\ and \psrb, very likely related to passage of the compact object through the dense equatorial wind of the B{\em e} companion (\ls\ has a O9V companion -- and no such outburst), the emission from the systems is surprisingly steady up to timescales of years. This contrasts with typical HMXBs. Their spectral energy distributions superpose well with each other, with a rising spectral luminosity from radio to 0.1-1~MeV, a flat spectrum up to $1-10$~GeV and a drop at TeV energies.

\subsection{What powers the VHE emission?}
The high-energy emission could be powered by accretion {\em via} capture of the intense stellar wind from the companion, but the Bondi-Hoyle rates are low and observational signatures of accretion are lacking. Still, microquasar-type models have been proposed in which the $\gamma$-rays are emitted in a relativistic jet (see contribution by Paredes in this volume). Resolved compact radio emission in \ls\ and \lsi\ on scales of 10-100 mas and interpreted as jet emission, constitutes a strong motivation.

On the other hand, the spectral and temporal similarities between the three binaries detected in VHE $\gamma$-rays hint at a common scenario. This is naturally provided for by \psrb. In this system, the relativistic wind emitted by the young 48~ms pulsar is contained by the stellar wind. Particle acceleration at the termination shock leads to X-ray synchrotron emission and $\gamma$-ray inverse Compton emission with the star photons. The measured pulsar spindown power is a modest $8\cdot 10^{35}$~erg~s$^{-1}$, implying rotation-power is converted into radiation with a high efficiency. 

The X-ray spectra have photon index $\Gamma\approx 1.5$, harder than
those of Crab-like high-luminosity pulsars but comparable to those of
pulsar wind nebula (PWN) powered by young pulsars with
$\dot{E}\sim10^{36}$~erg~s$^{-1}$ \citep{gotthelf}. X-ray pulses might
be detectable underneath the PWN emission if scattering is
negligible. However, pulsed radio emission would be strongly absorbed
by the dense stellar wind in \ls\ and \lsi, with free-free opacities
$\tau\ga 10^3$. The pulsed radio emission does vanish in \psrb\ around
periastron, when the pulsar probes the densest region of the Be
wind. At that point, the orbital separation is comparable to the
maximum separation in \lsi, and much larger than in the very compact
\ls.

The advantages of the PWN scenario, discussed below and in \citet{dubusp}, are that it readily explains the spectrum, level and stability of the emission without requiring additional assumptions from what is already known from plerions, and that it offers a very fruitful, common framework for interpretation.

\section{High energy emission from a compact PWN}
After the termination shock, a nebula of accelerated particles forms
behind the pulsar, in the direction opposite to the apparent
motion. Numerical simulations show the flow can be quite complex, but
the major factor in setting the properties of the nebula is the
distance of the termination shock $R_S$ from the pulsar. $R_S$ is set
by the balance of the ram pressures of the stellar and pulsar winds.

\subsection{The model}
A straightforward approach to test a PWN scenario for \ls\ and \lsi\ is to assume a pulsar with properties similar to those of \psrb\ but to change the stellar wind and orbit according to the optical observations of the binaries. The pulsar wind is therefore assumed to carry an energy $\approx 10^{36}$~erg~s$^{-1}$, composed of mono-energetic $e^+e^-$ pairs with Lorentz factor $\gamma_w \approx 10^5$ and a small magnetic field so that the ratio of magnetic to kinetic energy density $\sigma \approx 10^{-2}$, all reasonable values for a young, spinning-down pulsar.

For LS~5039, the wind mass-loss rate is $\approx 10^{7}$~M$_\odot$~yr$^{-1}$ with a terminal velocity $\approx 2000$~km~s$^{-1}$ so that 
\begin{equation}
\dot{E}/4\pi R_s^2 c = \rho_w v_w^2 ~~\rightarrow~~R_s\approx 1.5 \cdot 10^{11} d_{0.1}\dot{E}_{36}^{1/2}~{\rm cm}
\end{equation}
where $d_{0.1}=0.1$~AU is the orbital separation. The termination shock occurs close to the pulsar at a small fraction of the orbital separation.

The principles of an emission model are easily derived by analogy with plerionic supernova remnants. MHD shock conditions set the magnetic field intensity, flow speed ($\approx c/3$), density etc. Particles are assumed to be accelerated to a $dN\propto \gamma^{-2}d\gamma$ power-law up to the energy at which radiative losses occur on a timescale smaller than Bohm diffusion. Density conservation then sets the minimum $\gamma$ of the distribution. The synchrotron and inverse Compton emission on the stellar photon field (which occurs in the Klein-Nishina regime) can then be calculated. Particles cool as they move away from the pulsar; changes in flow conditions are given by a Bernouilli equation.

\subsection{Spectral energy distribution}
The major features of the spectral energy distribution derive from simple considerations on timescales. The acceleration timescale for TeV electrons ($\gamma_6$=$10^6$) is $t_{\rm acc}\approx 0.06~ \gamma_6 / B_1$~s with $B_1$=1~G the typical magnetic field intensity. The synchrotron cooling timescale is $t_{\rm S}\approx 770 / B_1^2 \gamma_6 {\rm ~s}$. Inverse Compton losses on stellar photons occur on a timescale $t_{\rm IC}\approx 20~ \gamma_6 d_{0.1}^2 / \left[\ln \gamma_6+1.4\right] (T_{\star, 4}R_{\star, 10})~{\rm s}$ with $T_\star$=40,000~K and $R_\star$=10~$R_\odot$ appropriate for the O star in \ls.

At the highest energies, $\gamma_{\rm max}$ is set by synchrotron losses ($t_{\rm acc}$=$t_{\rm S}$), which dominate over IC losses above a critical $\gamma_{\rm brk}$ given by ($t_{\rm S}$=$t_{\rm IC}$):
\begin{equation}
\gamma_{\rm brk}\approx 6 \cdot 10^6 ~(T_{\star, 4}R_{\star, 10})/ (B_1 d_{0.1}).
\end{equation}
Assuming continuous injection of electrons with $\gamma^{-2}$ spectrum, the steady-state distribution is steepened by radiative losses above $\gamma_{\rm brk}$ to $\gamma^{-3}$. This produces a flat synchrotron spectrum (in $\nu F_\nu$) above the frequency $\nu_{\rm S}$ corresponding to $\gamma_{\rm brk}$ (Eq.~3). The inverse Compton spectrum in Klein-Nishina regime roughly reflects the particle distribution with $\nu F_\nu \propto \nu^{-1.5}$ \citep{moderski} above $h\nu_{\rm IC}$=$\gamma_{\rm brk}m_ec^2$ (Eq.~4). Inefficient Klein-Nishina losses dominate below $\gamma_{\rm brk}$, producing a hard injection-like spectrum. Hence, the synchrotron spectrum below $\nu_{\rm S}$ is $\nu F_\nu \propto \nu^{0.5}$ and the inverse Compton spectrum below $\nu_{\rm IC}$ is flat. The characteristic frequencies are given by
\begin{eqnarray}
h\nu_{\rm S}& \approx &750 \left(T_{\star, 4}R_{\star, 10}/d_{0.1}\right)^2 /B_1 ~{\rm keV}\\
h\nu_{\rm IC}&\approx & 4 \left(T_{\star, 4}R_{\star, 10}/d_{0.1}\right)/{B_1} ~{\rm TeV.}
\end{eqnarray}

These considerations are quite general and will apply to any leptonic model close to the compact object. Numerical SEDs for the three binaries are shown in Fig.~1, consistent with the above expectations. The radio to MeV fluxes are in good agreement with observations. EGRET fluxes are underestimated but uncertainties in the GeV Galactic background emission could explain the discrepancy (cascades too). The TeV is discussed below.

\begin{figure}[t]
\centering\resizebox{6.25cm}{!}{\includegraphics{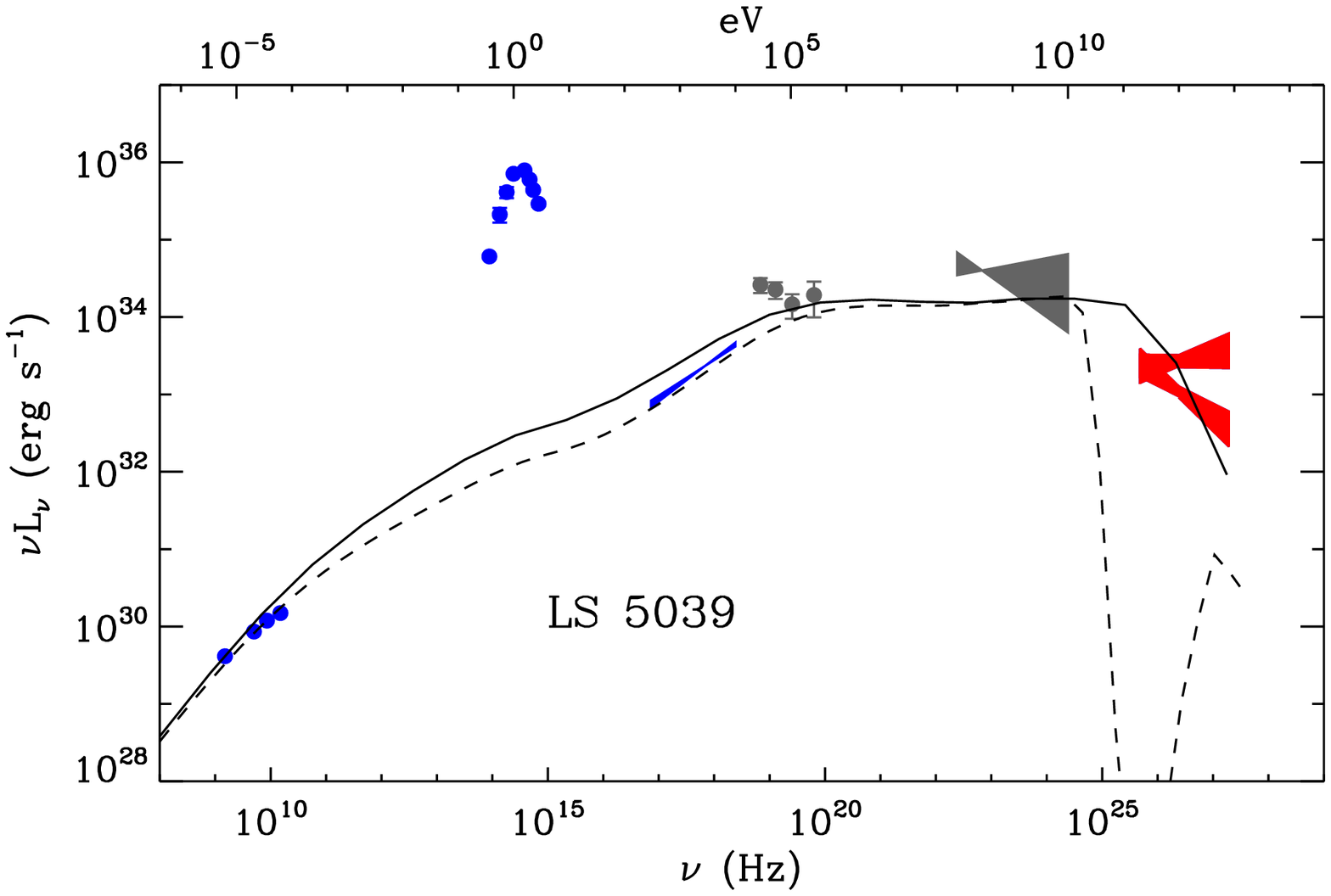}}
\centering\resizebox{6.25cm}{!}{\includegraphics{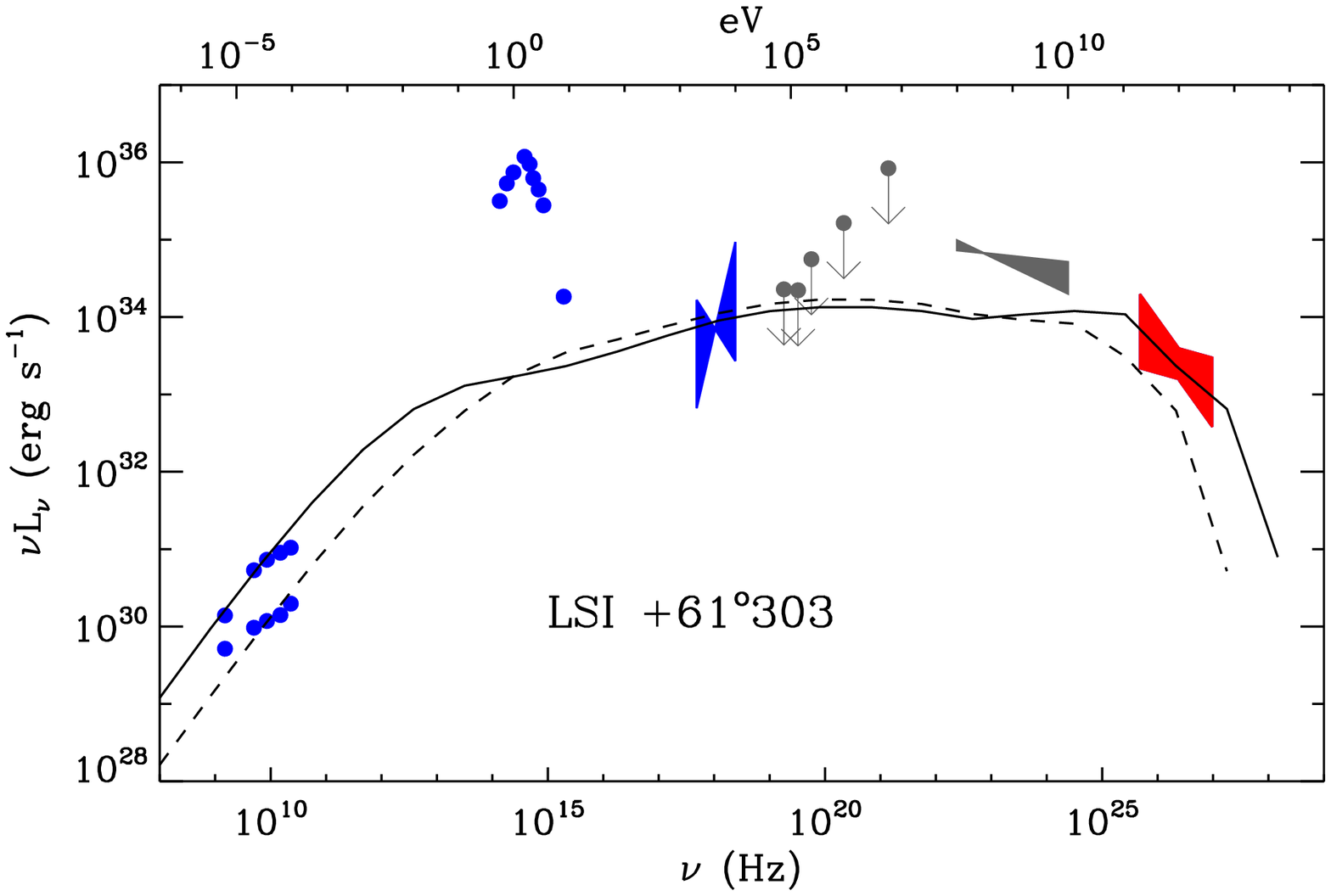}}
\centering\resizebox{6.25cm}{!}{\includegraphics{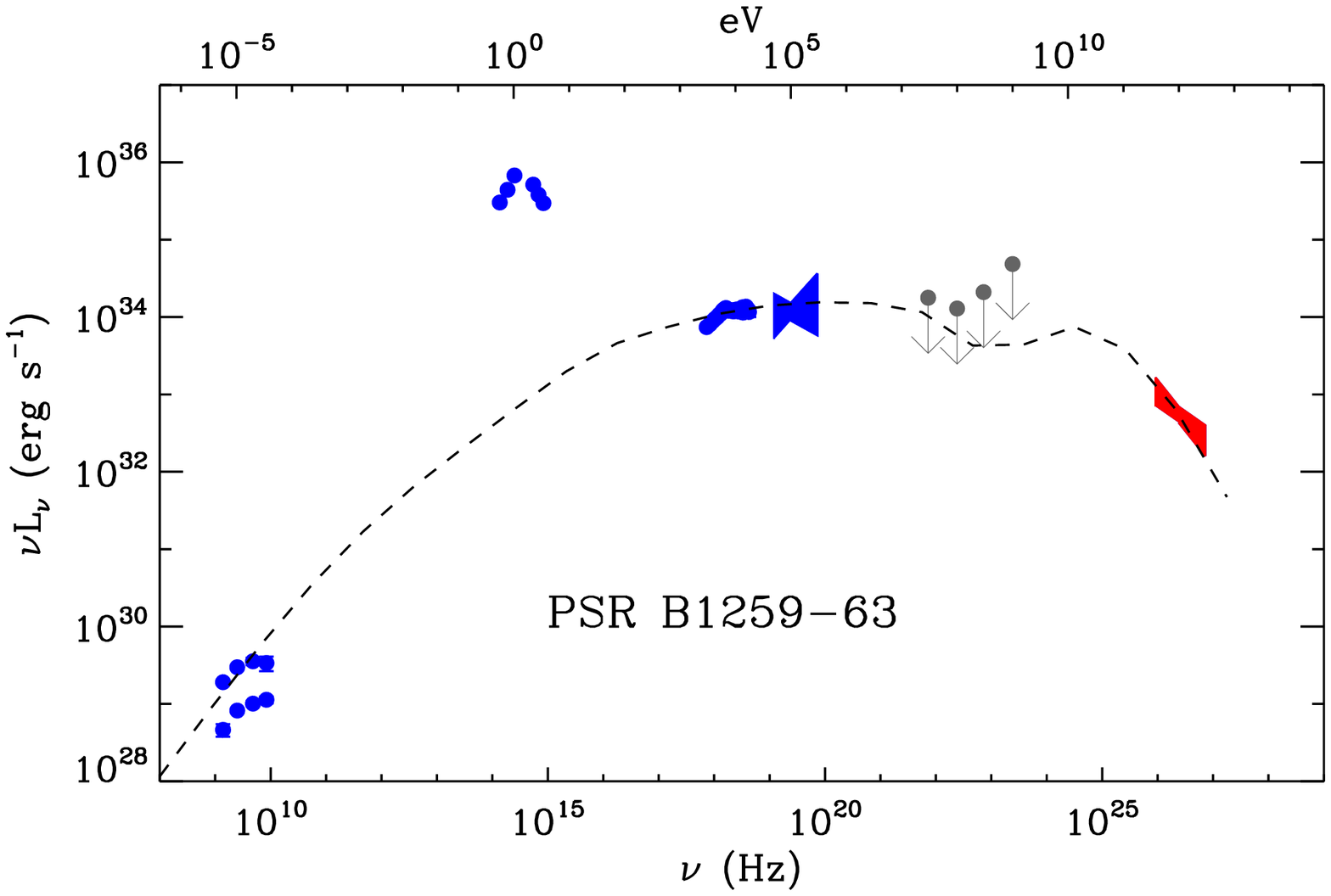}}
\caption{Spectral energy distributions for the $\gamma$-ray binaries with PWN emission models. For \ls, the pulsar wind has $\dot{E}$=10$^{36}$ erg/s, $\gamma_w$=5 10$^4$ and $\sigma$=0.01. The termination shock is at $R_s$=4 10$^{11}$~cm at apastron (solid line) and 2 10$^{11}$~cm at periastron (dashed line). The plot includes VHE $\gamma\gamma$ absorption (Fig. 2) but does not include emission from an associated cascade. For \lsi,  $\dot{E}$=10$^{36}$ erg/s, $\gamma_w$=5 10$^4$, $R_s$=2 10$^{11}$~cm and $\sigma$=0.02 at periastron in the dense equatorial wind (dashed line), changing to $R_s$=4 10$^{12}$~cm and $\sigma$=0.005 at apastron in the polar wind (solid line). The small $R_s$ at periastron implies a high $B$ so enhanced X-ray synchrotron and lower IC at VHE energies. For \psrb, the periastron parameters (dashed line) are $R_s$=10$^{12}$~cm, $\dot{E}$=8 10$^{35}$ erg/s, $\gamma_w$=4 10$^5$ and $\sigma$=0.005.}
\label{fig02} 
\end{figure}

\subsection{Periodicity in \ls}
An attractive feature of the PWN model is that the evolution of the SED with orbital phase is very easy to understand. If the stellar wind does not vary along the orbit, the magnetic field intensity is tied to $R_S$ so that  $B\propto 1/R_s \propto 1/d$ (Eq.~1). Hence, the spectrum should be unchanged at TeV energies but the hard X-ray spectral break should move as $1/d$. 

In \ls, $d$ varies only by a factor 2, so few changes are expected in the intrinsic SED. However, TeV photons embedded deep in the stellar radiation field can pair produce with the UV photons, leading to periodic absorption with maximum flux at inferior conjunction ($\phi$=0.7, Fig.~2), when the observed TeV photons travel directly away from the star \citep{dubus}. A few weeks after giving this talk, HESS reported that the TeV flux is indeed strongly modulated on $P_{\rm orb}$, confirming this basic picture and excluding emission far down ($>$1 AU) a relativistic jet \citep{mathieu}. The modulation is stable, as expected in the PWN model.

Spectral changes are observed with the modulation (Fig.~1). At low fluxes ($\sim$ periastron), the spectrum is steep. The source is never fully absorbed: VHE absorption must initiate an electro-magnetic cascade re-emitting $\gamma$-rays at lower energies, notably in the EGRET band. Obviously, this complicates the interpretation. At high fluxes ($\sim$ apastron where there should be little absorption), the TeV spectrum is flat with a break at $\approx 8$~TeV, suggesting $\gamma_{\rm brk}$ changes strongly with phase.  The geometry of the interaction region may play a crucial role here. For instance, a value of $B$=0.25~G (instead of 4~G in Fig.~1) would produce a 8~TeV break but implies a very low-$\sigma$ pulsar wind reaching to the star surface. Alternatively, an additional component might produce this hard spectrum, perhaps hadronic: theoretical investigations show nuclei in the pulsar wind are necessary for significant non-thermal acceleration.

\begin{figure}
{\resizebox{6cm}{!}{\includegraphics{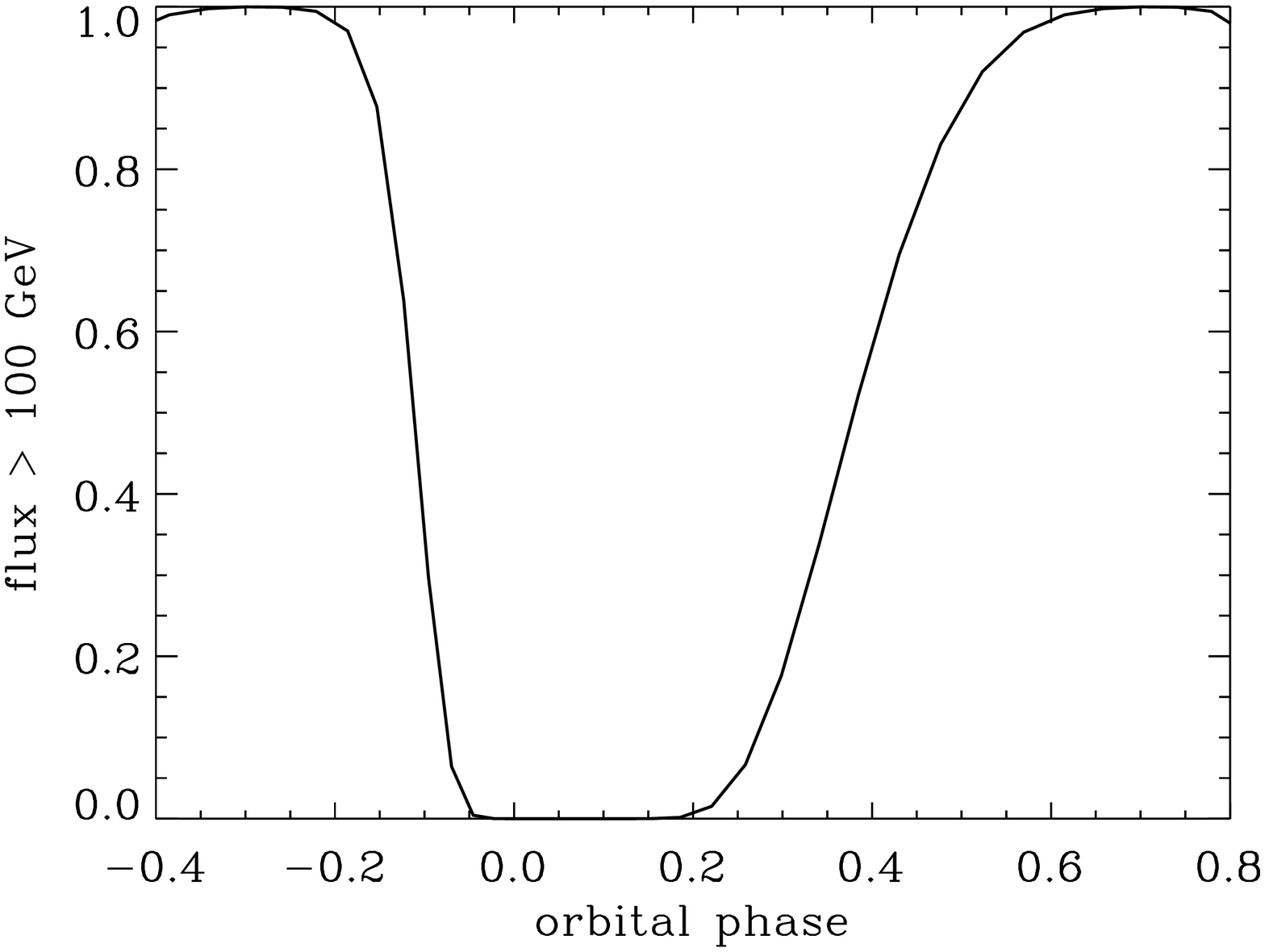}}}
{\resizebox{6cm}{!}{\includegraphics{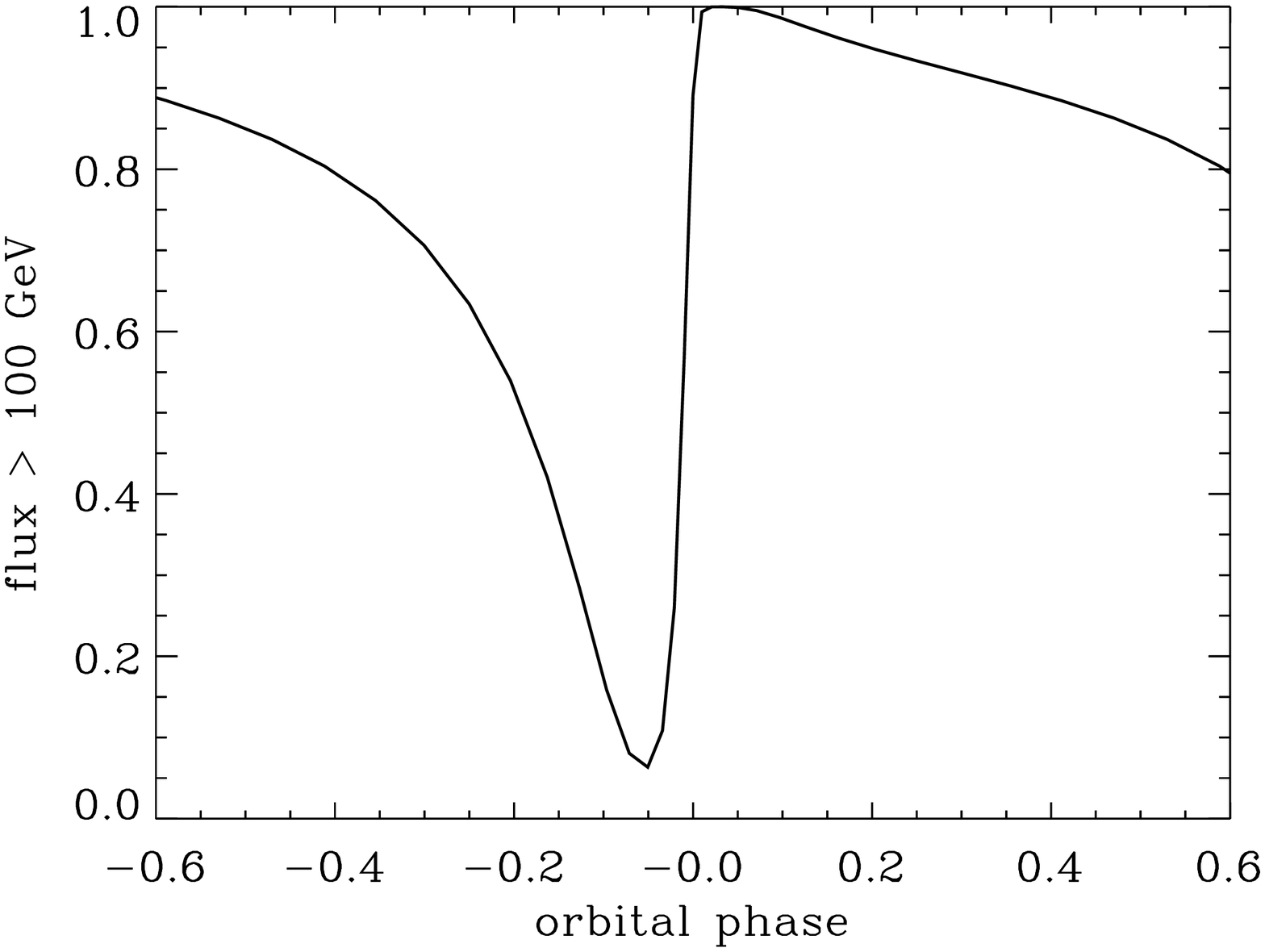}}}
\caption{Expected modulation of the integrated VHE flux above 100~GeV due to pair production on stellar photons in \ls\ (left) and \lsi\ (right). For \ls, peak emission is at $\phi$=0.7, as observed by HESS. Non-zero TeV fluxes are also detected at $\phi$=0-0.2 (Fig.~1), probably due to cascading. For \lsi, MAGIC detects the source at $\phi$=0.3-0.5 but sees no flux at phases 0-0.1 despite little absorption. The variability is more likely to be intrinsic. (Note that periastron is often defined set at $\phi$=0.23 instead of $\phi$=0 in \lsi.)}
\label{fig01} 
\end{figure}

\subsection{Periodicity in \lsi\ and \psrb}
MAGIC recently reported detecting \lsi\ at apastron but did not detect it at phases $\phi$=0-0.1 when the pulsar is in front of the star {\em i.e.} when absorption is actually {\em weakest}\footnote{\citet{boettcher} find $\gamma\gamma$ absorption explains the non-detection (at $\phi$=0-0.1) using $\phi$=0.93, {\em i.e.} superior conjunction where pair production is maximal.} (Fig.~2). In \lsi, the wider orbit and orientation are such that the opacity from the pulsar towards the observer is at most 2, while it reaches 40 in \ls.

The TeV variability in \lsi\ is more likely due to the presence of a {\em Be} star than to $\gamma\gamma$ absorption. Be stars have a slow, dense equatorial wind and a fast, polar wind (similar to that of the O star in \ls). The equatorial wind extends up to $\sim 20~R_\star$ so that the pulsar in \lsi\ and \psrb, who both have Be companions, plunges periodically through the dense material. This crushes the PWN, like the Earth magnetosphere during a Solar storm. The termination shock is much closer to the pulsar so $B$ is much higher, synchrotron losses are greater and the IC emission lower. If $B$ is high, $\nu_{\rm IC}$ is low and the steep IC spectrum leads to fainter TeV emission. At apastron, the stellar wind is tenuous, the termination shock further away from the pulsar, $B$ is low, $\gamma_{\rm brk}$ higher and IC emission more important at TeV, just as observed by MAGIC. This was predicted in \citet{dubusp}.

In \psrb\ the HESS lightcurve indicates a more complex, asymetric behaviour around periastron. An important thing to note when comparing the two systems is that the {\em periastron} separation in \psrb\ is close to the {\em apastron} separation in \lsi, and that the Be disk appears inclined with respect to the orbital plane in \psrb. Therefore, \psrb\ samples the equatorial disk {\em twice} around $\phi$=0 and at comparatively lower densities; IC emission would first decrease and then recover after each passage. This works well: the X-ray and TeV emission show peaks after the disc crossings inferred from radio dispersion measures\footnote{The peaks are less likely to be due to the actual Be disc crossing, as proposed by \citet{cnl}. The required orientation of the Be disc with respect to the orbit would imply the 2$^{\rm nd}$ crossing occurs 30 days after periastron, inconsistent with the pulsar radio eclipse {\em ceasing} 10 days earlier than that. Furthermore, the compact object is then 40~$R_\star$ away from the star, beyond the truncation radius of a typical Be disc (20~$R_\star$).}.

Further complications in computing the high energy spectra and lightcurves for the binaries may arise from IC anisotropy: the respective locations of the electrons, seed photons and observer can have an important effect, just as when computing $\gamma\gamma$ absorption \citep{kirk}. The high velocities of the shocked material ($c/3$ and higher in numerical simulations) can also cause Doppler (de)boosting effects. Finally, the pulsar wind itself is very likely anisotropic.

\section{Radio tails}
The radio emission in \lsi\ and \ls\ has been resolved on angular scales of 10-100 mas and widely interpreted as due to relativistic jet emission by analogy with compact jets in X-ray binaries \citep{paredes02,Massi2004}. The one-sidedness of the radio emission in \ls\ was interpreted as a Doppler effect, implying a jet speed of $\approx c/3$. Strong changes in the orientation of the (one-sided) radio emission in \lsi\ on a timescale $\ll P_{\rm orb}$ were interpreted as jet precession. On larger scales, the radio emission appears to be more double-sided in \ls\ and stable.

Emission from cooled PWN particles can explain the resolved radio fluxes very well. Isolated ms pulsars interacting with the ISM are known to emit well-collimated cometary tails of cooling shocked material in the direction opposite to motion, extending to parsec scales. Furthermore, colliding winds in some Wolf-Rayet binaries produce non-thermal radio emission whose appearance changes with orbital phase \citep{dougherty}. In a $\gamma$-ray binary, the appearance of the PWN will combine the effects of cooling in a comet tail whose direction is changing with the pulsar's orbital motion.

The aspect depends on orbital elements, shock geometry and cooling physics; but its basic features can be derived. For a face-on system ($i$=0$^{\rm o}$), the orbital motion creates a spiral of decreasing intensity with distance. The speed of the cometary material is typically $\sigma c$ so the step of the spiral is $\sim \sigma c P_{\rm orb}$. When the spiral is seen edge-on ($i$=90$^{\rm o}$), the PWN has a double-sided, jet-like aspect with peak intensity alternating on both sides with $P_{\rm orb}$. Inclination effects are less pronounced in a highly eccentric system, which has a clear preferred direction for most of the orbit, with dramatic changes in the position angle of the tail when the pulsar moves around periastron\footnote{This could explain the large P.A. changes seen by \citet{Massi2004} in \lsi\ ($e$=0.72) compared to \ls\ ($e$=0.35). Unfortunately, there are no high resolution maps of \psrb\ ($e$=0.87) available for comparison. Both radio PWN should be easier to resolve than in \ls\ as the orbital timescales are longer.}.

The PWN model predicts that the radio morphology at high resolution ($\sim \sigma c P_{\rm orb}$) should change on the orbital period. Maps at such resolutions depend mostly on geometric orbital effects and less on details of cooling. The milliarcsecond \ls\  map close to periastron is in excellent agreement with the published VLBI map (Fig.~3). At resolutions $\gg \sigma c P_{\rm orb}$ the appearance is due to the summed emission of particles over many orbits (spiral steps), so the emission will be stable in morphology and intensity. Good knowledge of particle cooling and flow conditions are then required to make robust predictions at low resolutions.

\begin{figure}
\centering\resizebox{5cm}{!}{\includegraphics{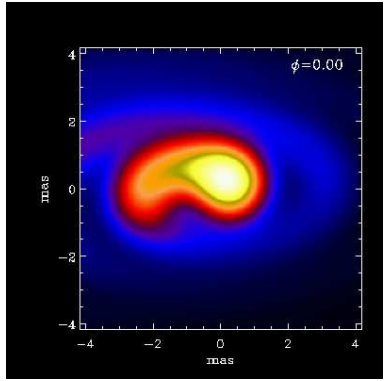}}
\caption{Radio appearance of the PWN nebula of \ls\ on milliarcsecond scales at $\phi$=0. The basic shape is a spiral of decreasing intensity with a step $\sigma P_{\rm orb}c$ ($\approx$2~mas here with $\sigma$=0.01). The projection on the sky assumes $i$=60$^{\rm o}$. The curved jet-like emission reproduces very well the only available VLBI observation. The emission will move left and right with orbital phase as the pulsar follows its orbit. At resolutions $\gg \sigma P_{\rm orb}c$, far-away particles from many orbits contribute, and the change in morphology is averaged out, leaving a steady radio source with an extension along the projected major axis.}
\label{fig04}
\end{figure}

\section{Conclusion}

Emission from a compact plerion provides a common, coherent framework to understand $\gamma$-ray binaries. These sources are the short-lived progenitors of accretion-powered HMXBs, which have slowly-rotating spundown pulsars. About 30 such sources are expected in our Galaxy\footnote{consistent with seeing 3 within 3~kpc} given the present-day HMXB population \citep{meurs89}. 

High-energy emission occurs on small scales. Strong variations in the location of the pulsar wind termination shock (\lsi, \psrb) and $\gamma\gamma$ absorption of VHE photons on starlight (\ls) can cause periodic changes at X-rays energies and above. Particles leave the vicinity of the pulsar as they cool, forming a cometary tail. Periodic changes in the aspect of this tail can be resolved in radio on scales $\sigma c P_{\rm orb}$, providing a test for the PWN scenario. The radio nebula is steady on larger scales.

Compact plerions enable access to pulsar wind physics on hitherto
inaccessible scales. Detailed understanding of the VHE emission,
notably in \ls, can provide important clues as to the shock geometry,
magnetic field intensity and acceleration processes around young
rotation-powered pulsars.











\end{document}